\documentclass[10pt]{IEEEtran}
\usepackage{bm}
\usepackage{epsfig}
\usepackage{algorithmic}
\usepackage{graphicx}
\usepackage{subcaption}
\usepackage{float}
\usepackage{cite}
\usepackage{url}
\usepackage{color}
\usepackage{balance}
\usepackage{mdwlist}
\usepackage{multirow}
\usepackage{threeparttable}
\usepackage{enumitem}
\usepackage{amsmath}
\usepackage{stmaryrd}
\usepackage{booktabs}
\usepackage{siunitx}
\usepackage[ruled,vlined]{algorithm2e}
\usepackage{threeparttable}
\usepackage{cuted}
\usepackage{graphicx}
\usepackage{epsfig,amsmath,amsfonts,cite}
\usepackage{listings}

\DeclareMathAlphabet\mathbfcal{OMS}{cmsy}{b}{n}
\newcommand{\ten}[1]{\mathbfcal{#1}}
\newcommand{\mat}[1]{\mathbf{#1}}

\def\sssp{\def\baselinestretch{0.88}\large\normalsize}\sssp

\begin{document}

\title{Tucker Tensor Decomposition on FPGA}

\author{
 Kaiqi Zhang, Xiyuan Zhang and Zheng Zhang\\

          {Department of Electrical \& Computer Engineering, University of California, Santa Barbara, CA 93106}\\
          {{kzhang70@ucsb.edu; xiyuanzhang@ucsb.edu; zhengzhang@ece.ucsb.edu}}
   }

\maketitle
\begin{abstract}
Tensor computation has emerged as a powerful mathematical tool for solving high-dimensional and/or extreme-scale problems in science and engineering. The last decade has witnessed tremendous advancement of tensor computation and its applications in machine learning and big data. However, its hardware optimization on resource-constrained devices remains an (almost) unexplored field. This paper presents an hardware accelerator for a classical tensor computation framework, Tucker decomposition. We study three modules of this architecture: tensor-times-matrix (TTM), matrix singular value decomposition (SVD), and tensor permutation, and implemented them on Xilinx FPGA for prototyping. In order to further reduce the computing time, a warm-start algorithm for the Jacobi iterations in SVD is proposed. A fixed-point simulator is used to evaluate the performance of our design. Some synthetic data sets and a real MRI data set are used to validate the design and evaluate its performance. We compare our work with state-of-the-art software toolboxes running on both CPU and GPU, and our work shows $2.16 -30.2\times$ speedup on the cardiac MRI data set.

\end{abstract}

\section{Introduction}
As a multi-dimension extension of matrices, tensors are a natural tool to represent and process multi-way data arrays~\cite{kolda2009tensor}. For instance, an MRI sequence with three spatial dimensions and a time dimension can be naturally represented as a 4-way tensor. The convolution layer in a neural network is also a tensor. Leveraging various tensor decompositions~\cite{hitchcock1927expression, tucker1966some,oseledets2011tensor}, many high-dimensional data mining~\cite{kolda2008scalable}, machine learning~\cite{hawkins2019bayesian,ding2017compact, chien2018tensor,novikov2015tensorizing,lebedev2014speeding,yang2017tensor} and EDA~\cite{zhang2015enabling,zhang2017big} problems have been solved efficiently without suffering from the curse of dimensionality. 

Due to their superior performance in processing high-volume data, tensors have emerged as a promising tool to enable real-time machine learning and data analysis. Recently, tensor algorithms have achieved great success in training and compressing deep neural networks~\cite{hawkins2019bayesian, ding2017compact, chien2018tensor,novikov2015tensorizing,lebedev2014speeding,yang2017tensor}. The resulting tensorized models consume tremendously less memory, run-time and power than the original deep neural network models. Tensor algorithms have also been successfully employed to accelerate medical image analysis~\cite{roohi2016dynamic}, anomaly detection~\cite{li2011robust} and speech recognition~\cite{saito2011one}. 

Despite the rapid progress of tensor algorithms, the hardware/algorithm co-design of tensor computation on resource-constrained platforms remains a new and (almost) unexplored field. Some tensor libraries~\cite{kaya2016high,smith2017sparse,li2015input} have been developed for high-performance platforms like clusters and super computers. However, little algorithm-architecture co-design targeting on power and cost-limited devices has been done, which has limited the application of tensor-based data analysis and machine learning on IoT and edge devices. Although many hardware accelerators are available for matrix and vector computations~\cite{amira2001accelerating, dou200564} and have been applied to machine learning \cite{zhu2003fpga,wang2017dlau,irick2008hardware} and signal processing~\cite{stanislaus2013low}, they cannot handle tensor data, because the underlying theory and numerical procedures are fundamentally different. It is inefficient and error-prone to process tensor data by matrix- or vector-computation accelerators. Therefore, resource-constrained hardware accelerators for tensor computation are highly desired.

This paper presents, for the first time, a hardware accelerator for one of the most important tensor algorithms: Tucker decomposition~\cite{tucker1966some}. Tucker decomposition is a high-order generalization of singular value decomposition (SVD) and principal component analysis (PCA), and it often achieves orders-of-magnitude higher data compression ratio than matrix compression algorithms on multi-way data. This method has been widely used in facial recognition~\cite{vasilescu2002multilinear}, signal processing~\cite{sidiropoulos2017tensor}, deep learning~\cite{kim2015compression} and data mining~\cite{kolda2008scalable}. Tucker decomposition is often implemented via the high-order orthogonal iteration (HOOI)~\cite{de2000best}. This algorithm involves some computation-intensive operations such as the tensor-times-matrix (TTM) and matrix SVD. Meanwhile, handling the huge amount of tensor data on FPGA or ASIC is a challenging task. 

The contributions of our paper are summarized below:
\begin{itemize}[leftmargin=*]
\item On the hardware side, we present an hardware architecture for Tucker decomposition. We describe the design and data communication of three units: TTM, SVD via Jacobi iterations, and tensor permutation/reshaping.
\item On the algorithm side, we propose a warm-start algorithm to reduce the cost of the Jacobi iterations.
\item We analyze the performance of our accelerator, implement it on a Xilinx FPGA, and show the implementation results.
\item We compare our FPGA accelerator with some state-of-the-art algorithms on both CPU and GPU, and demonstrate its application on an MRI compression task. Our accelerator shows up to $30.2\times$ speedup on the MRI data set.

\end{itemize}


\section{Algorithm Background}
\label{sec:preliminaries}
In this section, we introduce the necessary background of tensors and Tucker decomposition.
\subsection{Tensors and Basic Tensor Operations}
\label{subsec:tensor}
\textbf{Notations:} We use boldface lower-case letters (e.g., $\mat{x}$) to denote vectors, boldface capital letters (e.g. $\bf{X}$) to denote matrices, and boldface Euler script letters (e.g. $\ten{X}$) to denote tensors. 

A tensor $\mathbfcal{X}\in \mathbb{R}^{I_1\times I_2 \cdots \times I_d}$ is a multidimensional data array. Here $d$ is called the order or way of $\ten{X}$. An integer $k\in [1,d]$ can be used as the index of a specific dimension or mode with size $I_k$. An entry of a tensor can be specified by an index vector. For instance, the $(i_1, i_2, \dots i_d)$-th entry in tensor $\ten{X}$ is denoted by $x_{i_1, i_2, \dots i_d}$. Clearly, a vector and a matrix are order-1 and order-2 tensors, respectively. 
\begin{figure}[t]
    \centering
    \includegraphics[width=0.65\linewidth]{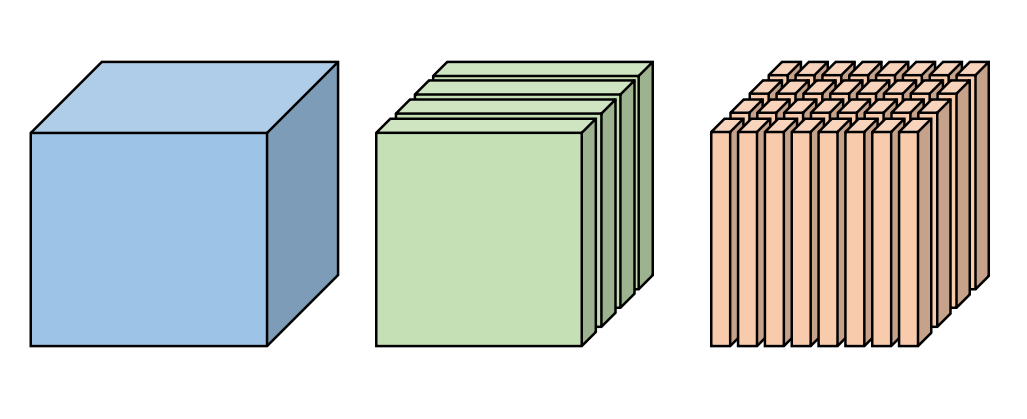}
    \caption{Left to right: a tensor, slices, and fibers.}
    \label{fig:fiber}
\end{figure}

\vspace{5pt}
\textbf{Tensor Fiber and Slice:} A \textbf{fiber} is a one-dimensional fragment of a tensor, obtained by fixing all indices but one. Tensor fibers are higher-order extension of matrix rows and columns. A third-order tensor has fibers that can be denoted by $\mat{x}_{:jk}$, $\mat{x}_{i:k}$ or $\mat{x}_{ij:}$ correspondingly.  A tensor \textbf{slice} is a two-dimensional fragment of a tensor, obtained by fixing all indices but two. For instance, $\mat{X}_{i::}$, $\mat{X}_{:j:}$ and $\mat{X}_{::k}$ denote the horizontal, lateral, and frontal slices of a $3$-way tensor, respectively. Fig.~\ref{fig:fiber} shows the slices and fibers of a tensor.

\vspace{5pt}
\textbf{Tensor permutation:}
Tensor permutation changes the mode order of a tensor. It is a high-order extension of the matrix transpose. For instance, given $\ten{X}\in \mathbb{R}^{5\times 10 \times 3}$, ${\bf permute}(\ten{X}, [2,3,1])$ generates a new tensor $\ten{Y}\in \mathbb{R}^{10\times 3 \times 5}$ with $y_{i_2,i_3,i_1}=x_{i_1,i_2,i_3}$.


\begin{figure}[t]
    \centering
    \includegraphics[width=0.75\linewidth]{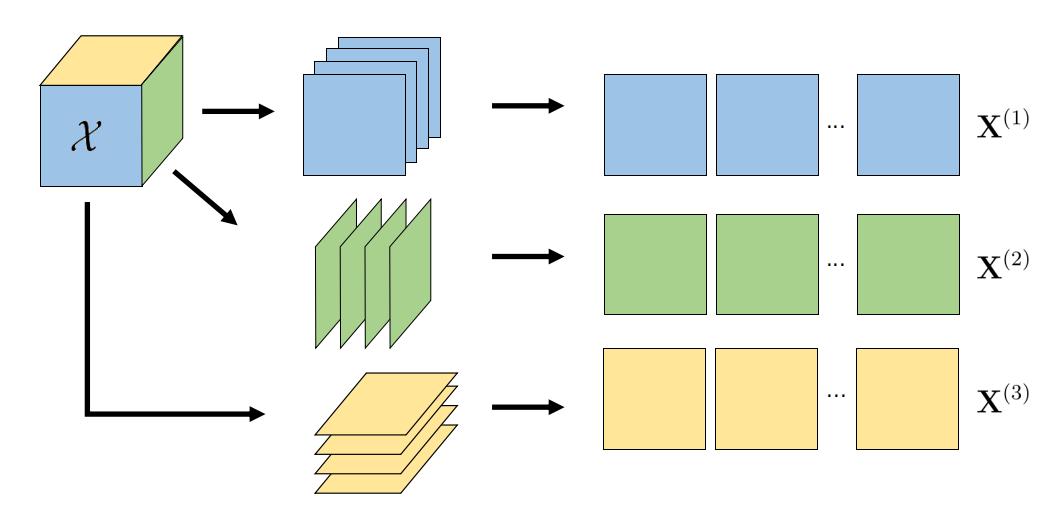}
    \caption{Tensor unfolding.}
    \label{fig:unfold}
\end{figure}

\vspace{5pt}
\textbf{Unfolding:} Unfolding (or matricization) as shown in Fig.~\ref{fig:unfold}, is the process of transforming a tensor into a matrix. Unfolding reorders the elements of an $d$-way data array into a matrix. For instance, the mode-$k$ unfolding of a tensor $\ten{X}\in \mathbb{R}^{I_1 \times I_2 \cdots \times I_d}$ is denoted as $\mat{X}^{(k)} \in \mathbb{R}^{I_k \times \prod\limits_{m\neq k} I_{m}}$. The mapping from the $(i_1,\dots,i_d)$-th element $\ten{X}$ to the $(i_k,j)$-th element of $\mat{X}^{(k)}$ is given as follows
\begin{equation}
    j=1+
    \sum_{m=1,m\neq k}^{d}\left({(i_{m}-1)}\prod_{n=1,n\neq k}^{m-1}{I_n} \right).
\end{equation}

\vspace{5pt}
\textbf{TTM:} Tensor-times-matrix (TTM) in short, is a high-dimensional expansion of matrix multiplication. Given a tensor $\ten{X}\in \mathbb{R}^{I_1\times I_2 \cdots \times I_d}$ and a matrix $\mat{A} \in \mathbb{R}^{J\times I_k}$, their $k$-mode product is a new tensor
\begin{equation}
    \ten{Y} = \ten{X} \times_k \mat{A} \in \mathbb{R}^{I_1 \times \cdots I_{k-1}\times J \times I_{k+1} \cdots \times I_d}.
\end{equation}
It can be expressed as matrix-matrix multiplication
\begin{equation}
    \mat{Y}^{(k)}=\mat{A}\mat{X}^{(k)}
\end{equation}
However, implementing it in this way directly can be inefficient on hardware because tensor permutation will be needed. This needs to move almost all the data in a tensor.

\vspace{5pt}
\textbf{Tensor Norms:} The norm of a tensor is a function which maps any tensor to a non-negative scalar. A widely used norm of tensors is the Frobenius norm, defined as
\begin{equation}
    ||\ten{X}||_{\rm F}=\sqrt{\sum_{i_1, i_2, \dots i_d}{x}_{i_1, i_2, \dots i_d}^2}.
\end{equation}

\subsection{Tenser Tucker Decomposition}
\label{subsec:tucker}

\begin{figure}[t]
    \centering
      \includegraphics[width=0.8\linewidth]{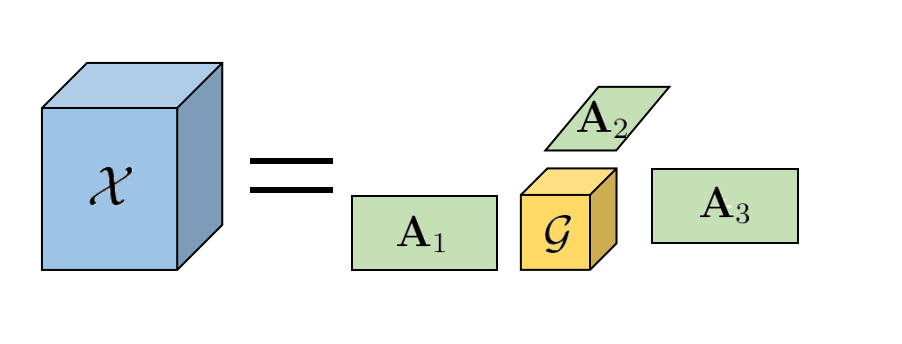}
    \caption{Tucker decomposition.}
    \label{fig:tucker}
\end{figure}

Given an $d$-way tensor $\ten{X} \in \mathbb{R}^{I_1 \times I_2 \times \dots \times I_d}$, we may compress it by the Tucker decomposition~\cite{tucker1963implications}. As shown in Fig.~\ref{fig:tucker}, the Tucker decomposition approximates a large-size tensor with a small-size core tensor $\ten{G} \in \mathbb{R}^{R_1 \times R_2 \cdots \times R_d}$ and $d$ factor matrices $\{ \mat{A}_{k} \in \mathbb{R}^{I_k \times R_k}\}_{k=1}^d$ as follows:
\begin{equation}
    \ten{X} \approx \ten{G} \times_1 \mat{A}_1 \times_2 \mat{A}_2 \times \dots \times_d \mat{A}_d, 
\end{equation}
where all columns are orthonormal inside each factor matrix $\mat{A}_k$. 
We call $\mat{R}=[R_1, R_2, \cdots, R_d]$ as the \textbf{multilinear rank} of $\ten{X}$. When $R_k\ll I_k$, $\ten{X}$ can be represented with the above Tucker format at a much lower cost. Once all factor matrices are obtained, the core tensor can be computed as
\begin{equation}
    \ten{G} = \ten{X} \times_1 \mat{A}_1^T \times_2 \mat{A}_2^T \times \dots \times_d \mat{A}_d^T. 
\end{equation}

Two popular methods can be used to compute a Tucker decomposition. The first well-known method is the high-order SVD (HOSVD) \cite{tucker1966some}. The idea of HOSVD is simple: 
\begin{itemize}[leftmargin=*]
    \item  One first unfolds $\ten{X}$ along mode $k$ to get $\mat{X}^{(k)}$;
    \item Then, perform a singular value decomposition (SVD) 
    \begin{equation}
        \mat{X}^{(k)}=\mat{U}_k \mat{S}_k \mat{V}_k^T.
    \end{equation}
    
    \item Finally, pick $\mat{A}_k$ as the first $R_k$ columns of $\mat{U}_k$.
    
\end{itemize} 
This method is easy to implement, however it is not optimal in fitting the data. Alternatively, an alternative least-square method called HOOI is widely used to get a better solution.  

\vspace{5pt}
\textbf{HOOI:} The High Order Orthogonal Iteration (HOOI) \cite{de2000best, kroonenberg1980principal, kapteyn1986approach} method aims to minimize the approximation error
\begin{equation}
    \min_{\{\mat{A}_k \}_{k=1}^d } || \ten{X} - \ten{G} \times_1 \mat{A}_1 \times_2 \mat{A}_2 \times \dots \times_d \mat{A}_d||_F
\end{equation}
through the iterative process as shown in Alg.~\ref{alg:hooi}. Each iteration of the HOOI consists of two steps for every mode index $k$: (1) obtain a tensor $\ten{B}$ via a power iteration (TTM along all modes except $k$), which can be done from mode $d$, then $d-1$, ... until mode 1. (2) an SVD of the mode-$k$ unfolded matrix of $\ten{B}$ to extract a mode-$k$ factor matrix $\mat{A}_k$.

In practice, the initialization process via HOSVD can be very time-consuming, because it needs $d$ SVD operations, and each of it works on a matrix whose size is equal to the original tensor. Therefore, some random orthonormal matrices are often used as the initial factor matrices. In this case, the total number of iterations needed may increase slightly, but the total runtime can decrease significantly. Even though, when the size of the tensor $\ten{X}$ is large, the time and energy comsumed to compute the Tucker decomposition can be very high.

\begin{algorithm}[t]
\caption{HOOI for Tucker Decomposition}
\label{alg:hooi}
\begin{algorithmic}[1]
\STATE { Initialize $\{\mat{A}_k\} _{k=1}^d$ via HOSVD}
\WHILE {not converge} 
    \FOR {$k=1,2, \ldots, d$} 
        \STATE {$\ten{B} \leftarrow \ten{X} \times_1 \mat{A}_1^T \dots \times_{k-1} \mat{A}_{k-1}^T \times_{k+1} \mat{A}_{k+1}^T \dots \times_d \mat{A}_d^T$ }
        \STATE {Unfold $\ten{B}$ and perform SVD: $ \mat{B}^{(k)}=\mat{U}_k \mat{S}_k \mat{V}_k^T $}
        \STATE {$\mat{A}_k \leftarrow$ the first $R_k$ columns of $\mat{U}_k$.}
    \ENDFOR
\ENDWHILE
\STATE {\textbf{return} $\{ \mat{A}_{k}\}_{k=1}^d$. }

\end{algorithmic}
\end{algorithm}

\section{Proposed Architecture}
\label{sec:architecture}

\begin{figure}[t]
    \centering
     \includegraphics[width=0.4\textwidth]{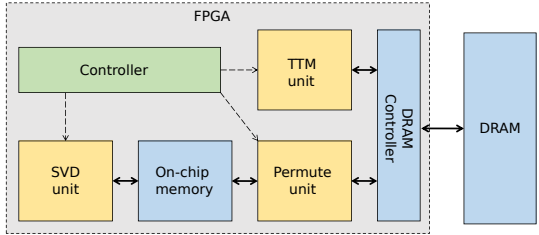}
    \caption{Overall structure of our Tucker decomposition.}
    \label{fig:blk}
\end{figure}

In this section, we propose a new hardware architecture to perform Tucker decomposition via HOOI. 

Fig. \ref{fig:blk} shows the system-level diagram of our architecture. In order to load and accommodate the huge amount of tensor data elements, our design stores the tensor in an external DRAM. The HOOI engine consists of three parts: a tensor-times-matrix (TTM) unit, a singular value decomposition (SVD) unit, and a tensor permuting/reshaping unit. The data elements of the three units are stored in different memories. Because the size of a tensor can be very large, all tensors (including the intermediate and final results of TTM) are stored in an external DRAM. The matrix for each SVD is stored in an on-chip memory to reduce latency and to achieve a maximum throughput. Both parts are parallel and pipelined to achieve the maximum performance. The tensor permuting unit moves the data between the on-chip memory used by the SVD unit and the external DRAM used by the TTM unit, and it permutes the tensor accordingly. 

\subsection{Tensor-Times-Matrix (TTM) Unit}
The TTM unit can be implemented as a matrix-matrix multiplication that is available in some common computational libraries. However, we need to permute tensors before it can be implemented in this way. This is time- and memory-consuming because almost all $\prod\limits_{k=1}^d I_k$ data elements of $\ten{X}$ have to be moved. In this work, we develop a TTM unit {\it without tensor permutations}. For simplicity, the tensors are always stored by incrementing the mode-1 index $i_1$, then the second index $i_2$, and so forth. Since the size of tensor $\ten{X}$ is often beyond the capacity of an on-chip memory, all data elements (including the input and output data, and the intermediate result of a TTM operation) are stored in an external DRAM.

\begin{figure}[t]
\centering
       \includegraphics[width=0.4\textwidth]{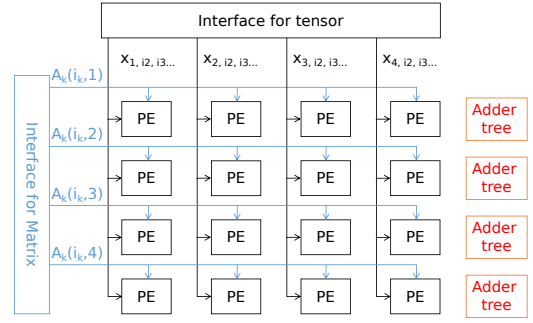}
  \caption{The TTM unit. The red part is used when computing the mode-1 product and blue part is used when computing mode-$j$ product when $j \neq 1$.}
    \label{fig:ttm}
\end{figure}

\vspace{5pt}
\textbf{TTM with a 2-D PE Array.} Our TTM unit is shown in Fig.~\ref{fig:ttm}. In order to maximize the throughput of the TTM module, we design a 2-D array of processing elements (PE) with $q$ columns and $r$ rows.  Each PE computes the product of one scalar element of the tensor (black arrow) and one scalar from the matrix (either from the blue line or stored in the PE) at each clock cycle. The PEs in a single column are always connected to the same bus so they handle the same element in the tensor $\ten{X'}$ each time. Here $\ten{X}'$ can be either the original tensor $\ten{X}$ or an intermediate tensor after the TTM of $\ten{X}$ with some factor matrices. 
With $q$ columns, this module can handle up to $q$ neighbouring elements of a tensor fiber in total at the same time. Due to our method of storing $\ten{X'}$, the fibers are always obtained by only changing the first mode index. 
Each row of this array handles one column of the factor matrix $\mat{A}_j$.

We compute the mode-$j$ TTMs in a decreasing order from $j=d$ to $j=1$. This PE array works in different ways dependent on the value of $j$, which is explained below. 
\begin{itemize}[leftmargin=*]
    \item Assume that $j\neq 1$ and that we have done TTMs for all modes $>j$ except mode $k$. Let us ignore mode $k$ for simplicity, and the size of $\ten{X'}$ becomes $I_1 \times I_2 \times \dots I_j \times R_{j+1} \times \dots \times R_d$. 
To simplify the expression, we fold the modes $1, 2, \dots j-1$ into a single mode and use $\hat{i}$ as its index. The element-wise expression of this operation is 
\begin{equation}
    y_{\hat{i}, r_j, i_{j+1}\dots i_d}=\sum_j^{I_j} x'_{\hat{i}, i_j, i_{j+1}\dots i_d}\mat{A}_j(i_j, r_j).
\end{equation}

In each clock cycle, each vertical bus can carry some neighbouring elements in a tensor fiber $\mat{x}'_{: i_j  \dots i_d}$ and each horizontal bus can carry an element in the factor matrix. Specifically, In the $n$-th cycle of the $i_j$-th round, the $l$-th vertical bus and the $r_j$-th horizontal bus carry scalars ${x'}_{\hat{i}, i_j,  \dots, i_d}$ and $\mat{A}_j(i_j, r_j)$, respectively. Consequently the $(r_j,l)$-th PE calculates
\begin{equation*}
    {x'}_{\hat{i}, i_j,  \dots, i_d}\times a_{i_j, r_j}, \;{\rm with}\; \hat{i}= l+nq\leq \hat{I}'=I_1I_2\cdots I_{j-1}.
\end{equation*}
Finally $y_{\hat{i}, r_j, i_{j+1},\dots, d}$ is obtained by summing the above product terms over $i_j$ (in each round). Note that the index of matrix elements in $\mat{A}_j$ used at each PE does not depend on $l$. Therefore, we can multiply the whole fiber $\mat{x}'_{: i_j  \dots i_d}$ with the same matrix element, and all PEs in the same row share the same data elements of $\mat{A}_j$ by connecting them to the same horizontal bus. Because the size the new dimension, $\hat{I}=I_1I_2\dots I_{k-1}$, is very large, we divide the partly-folded tensor $\ten{X}'$ into some small sub-tensors of size $m \times I_j \times R_{j+1}\dots \times R_d$, and the resulting tensor $\ten{Y}$ into some sub-tensors of size $m \times R_j \times R_{j+1} \dots \times R_d$. We can compute the sub-tensors of $\ten{Y}$ one by one to reduce the buffer size. 

\item When computing $\ten{X'} \times_1 \mat{A}_1^T$, the element-wise result is 
\begin{equation}
\label{ttm_mode_1}
    y_{r_1, i_2,  \dots, i_d} = \sum_{i_1 = 1}^{I_1} x'_{i_1, i_2, \dots, i_d} \mat{A}_1(i_1, r_1).
\end{equation} 
The $r_1$-th row of the PE array computes $q$ product terms of~\eqref{ttm_mode_1}. Because $q$ is usually smaller than the fiber length, we need to compute all product terms by several cycles. In the $n$-th cycle, the $(r_1, l)$-th PE calculates \begin{equation*}
    {x'}_{i_1, i_2,  \dots, i_d}\times a_{i_1, r_1}, \;{\rm with}\; i_1= l+nq\leq I_1.
\end{equation*}
Note that in mode-1 TTM,  the RAM inside the $(r_1,l)$-th PE stores all $a_{i_1, r_1}$'s for all $i_1=l+nq\leq I_1$. The TTM element $y_{r_1, i_2, \dots i_d}$ is obtained by 
accumulating all products terms in the $r_1$-th row during all cycles.
\end{itemize}


\begin{figure}[t]
\centering
 \begin{subfigure}[b]{0.2\textwidth}
    \centering
    \includegraphics[width=\textwidth]{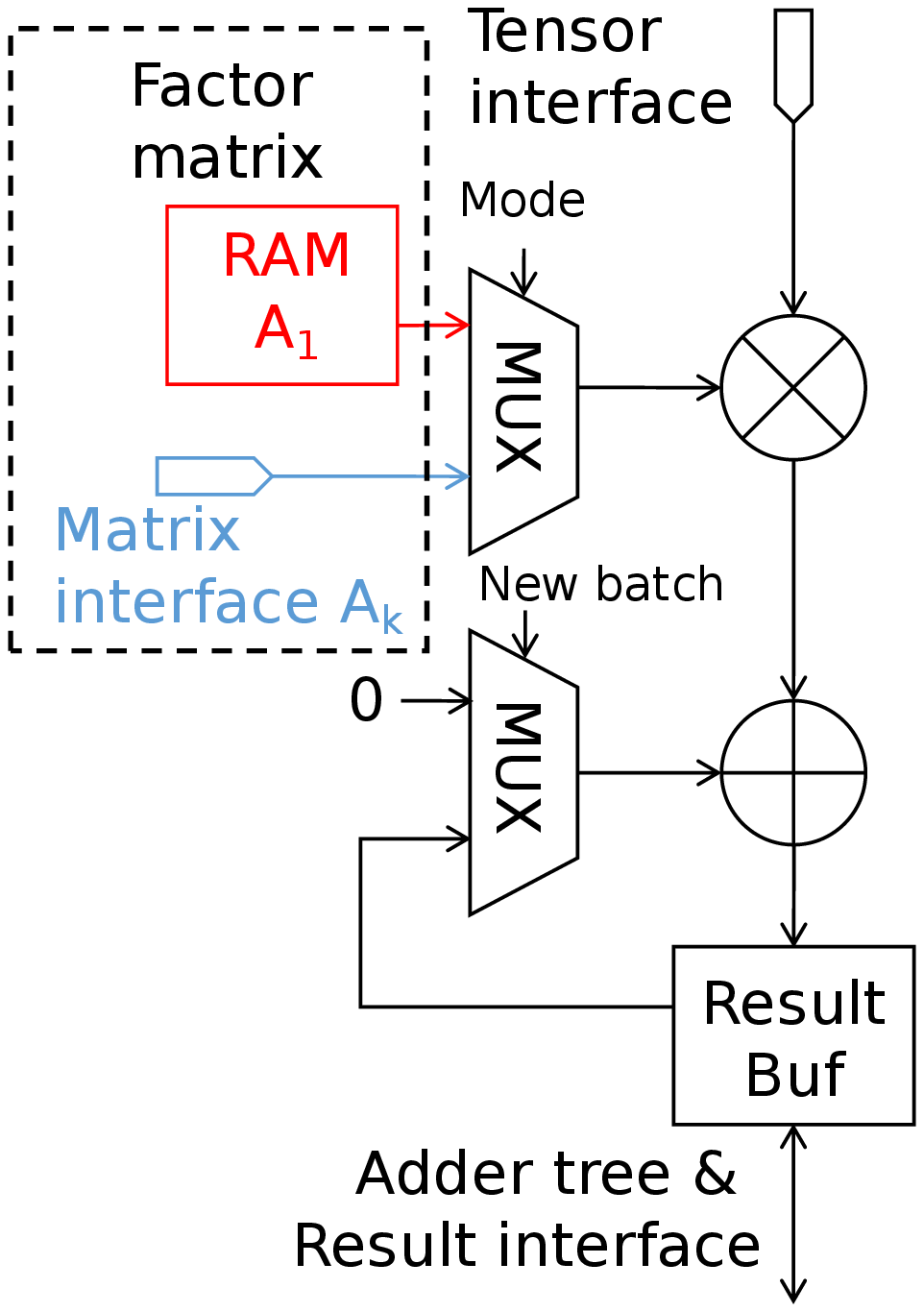}
    \caption{}
\end{subfigure} 
\begin{subfigure}[b]{0.2\textwidth}
        \centering
        \renewcommand{\lstlistingname}{}
        In-place adder tree:
        \small
       \begin{algorithmic}[1]
    \FOR {$j$=1 \TO log(n)} 
        \FOR{$i$=1 \TO n/2}
            \STATE{$a_i\leftarrow a_{2i-1}+a_{2i}$}
        \ENDFOR
        \FOR{$j$=n/2+1 \TO n}
            \STATE{$a_i\leftarrow 0$}
        \ENDFOR
    \ENDFOR
\end{algorithmic}
        \vspace{1.0cm}
         \caption{}
\end{subfigure}
  \caption{(a) The details of a PE. The red part is used only for mode-1 TTM, and blue part is only for mode-$j$ TTM with $j \neq 1$. (b) An in-pace adder tree for $\mat{a} \in \mathbb{R}^n$.}
    \label{fig:PE}
\end{figure}

\vspace{5pt}
\textbf{Processing Element (PE).} As shown in Fig.~\ref{fig:PE}, each PE consists of a multiplier, a small RAM, and another memory used as an output buffer storing the result before writing it to a DRAM. The RAM (marked in red) is only used to store $\mat{A}_1$ when computing the mode-1 TTM. Otherwise, the bus at each row imports data elements of $\mat{A}_j$ when $j \neq 1$. Therefore, a mutiplexer (MUX) is used to select the correct data to perform product operations. After computing one batch of the data (a tensor fiber if $j=1$ and a tensor slice if $j \neq 1$), the result is written to the DRAM and then reset to zero. The buffer temporarily holds the intermediate results, and keeps updating during the multiplication and sum operations. The buffer stops updating when its data is written into an external memory. In order to avoid timing conflicts and to increase throughput, another buffer is used (not shown in the figure) for transferring data to an external memory. These two buffers switch their roles after processing each batch of data.

\vspace{5pt}
{\bf In-place Adder Tree.} As mentioned above, we need an adder tree for the mode-1 TTM. 
If the adder tree is implemented as a pipeline, a lot of adders and registers will be used. Given that the product terms need to be summed up only once per batch of data instead of per clock cycle, we only need an in-place adder tree. We split a adder tree into multiple stages. After each stage, each two elements are summed up so the total number of data elements is reduced by $50\%$. The registers and adders are shared among different stages. The data elements are read from and write back to the same group of registers after each clock cycle. This is why we call it an ``in-place'' adder tree. This treatment can reduce the number of adders and registers by $50\%$.

\subsection{Singular Value Decomposition (SVD) Unit}

\begin{algorithm}[t]
\caption{SVD via single-side Jacobi iteration}
\label{alg:jacobi}
\begin{algorithmic}[1]
\STATE {Input: $\mat{B}=\mat{B}^{(k)}$, initial guess $\mat{U}=\mat{I}$.}
\WHILE{Not converge}
 \FOR{any $(i,j), 1\leq i < j\leq n$, $i \neq j$}
    \STATE{$\alpha = \|\mat{b}_{i,:}\|_2^2$, $\beta = \|\mat{b}_{j:}\|_2^2$, $\gamma = \langle \mat{b}_{i:},\mat{b}_{j:} \rangle$}
    \STATE { $\theta = \arctan(\frac{2\gamma}{\beta-\alpha})$}
    \STATE {$\mat{b}_{i:} = \mat{b}_{i:}\cos\theta - \mat{b}_{j:}\sin\theta$, $\;$ 
    $\mat{b}_{j:} = \mat{b}_{i:}\sin\theta + \mat{b}_{j:}\cos\theta$}
    \STATE {$\mat{u}_{i:} = \mat{u}_{i:}\cos\theta - \mat{u}_{j:}\sin\theta$, $\;$ 
    $\mat{u}_{j:} = \mat{u}_{i:}\sin\theta + \mat{u}_{j:}\cos\theta$}
 \ENDFOR
\ENDWHILE
\RETURN{$\mat{U}$, with its $i$-th row being $\mat{u}_{i:}$.}
\end{algorithmic}
\end{algorithm}

Our SVD unit employs the Jacobi iterations \cite{hansen1963cyclic,demmel1992jacobi,hemkumar1992systolic, ahmedsaid2003improved, rahmati2008fpga}. Both single-side and double-side Jacobi iterations are widely used. We use the single-side version because of its higher accurate, ease to parallelize and less data dependency. The whole framework is summarized in Alg.~\ref{alg:jacobi}. Given a matrix $\mat{B}$, this algorithms computes its left singular vectors by orthogonizing every two rows (i.e., $\mat{b}_{i,:}$ and $\mat{b}_{j,:}$) of the matrix iteratively. The iterative process involves the norms and inner products of the row vectors and performing some rotations.

We can select the order of $(i,j)$ in different ways. A natural choice is to increment $j$ in the inner loop and increment $i$ in the outer loop, or vice versa. However, because of the data dependency between two adjacent operations, this choice makes it impossible to implement parallel or pipe-lined design. In order to overcome this issue, we employ the round-robin ordering suggested in \cite{brent1985solution}, which eliminates the data dependency between adjacent iterations. 
This method starts by dividing all indices into $n/2$ pairs $(1,2), (3,4), \dots (n-1, n)$ where $n$ is the total number of rows. After orthogonalizing all the pair of rows specified above, we generate new index pairs in this way: suppose the pair in the previous round is $(p, q)$, this pair index is updated in the next round as 
$$
\left\{
\begin{array}{ll}
(p+1, q-1)  & \text{if } q-p > 2,\\
(1, p+q)    & \text{if } q-p \leq 2\; \text{and}\; p+q\leq n,\\
(p+q+1-n,n) &\text{if } q-p \leq 2\; \text{and}\; n<p+q<2n-1,\\
(1, 2)      & \text{if } q=n\; \text{and}\; p=n-1.
\end{array}
\right.
$$

\begin{figure}[t]
    \centering
    \includegraphics[width=0.44\textwidth]{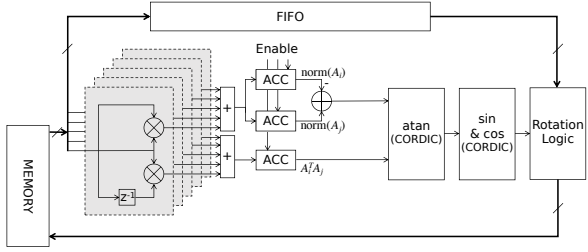}
    \caption{Block diagram of SVD unit. ACC: accumulator. FIFO: first-in, first-out queue.}
    \label{fig:svd}
\end{figure}

Fig.~\ref{fig:svd} shows the block diagram of our SVD unit. In each step, two vectors are orthogonalized. The on-chip memory provides two ports to operate independently. In this part, one port takes the two vectors from the memory,  and another port writes the orthogonalized vectors back into the memory. Because we use only one port to input data and the other for output, the two vectors have to be fetched through the same port alternatively. Three sum-of-products are needed to calculate the rotation angle $\theta$. Given that the two vectors are fetched alliteratively, the multiplier and adder to get $\|\mat{b}_{i:}\|_2$ and $\|\mat{b}_{j:}\|_2$ can be shared. Therefore, only two sets of multipliers and adders are used. In order to get $\theta$, $\sin\theta$ and $\cos\theta$, we employ the CORDIC algorithm \cite{volder1959cordic}, which uses the rotations of some fixed angles to approximate the rotation of any angle. This algorithm is efficient to calculate the trigonometric functions on hardware, and an FPGA IP core is available. After these two vectors are fetched from the memory, they are stored in a local buffer until $\sin\theta$ and $\cos\theta$ are calculated, then they are rotated accordingly. In this way, it is guaranteed each vector is read from the memory only once when orthogonalizing each pair of rows. 

Besides the input matrix $\mat{B}$, the orthogonal matrix $\mat{U}$ also needs to be rotated in the same way. We store $\mat{U}$ and $\mat{B}$ in the same memory and handle them in the same way, except that $\mat{U}$ is not used for calculating $\alpha, \beta$ and $\gamma$.
Since $\mat{U}$ has a much smaller size than $\mat{B}$, such a design only causes negligible run-time overhead but saves half of the area and power.

 \subsection{Tensor Permutation and Reshaping}
Once $\ten{B}=\ten{X} \times_1 \mat{A}_1^T \dots \times_{k-1} \mat{A}_{k-1}^T \times_{k+1} \mat{A}_{k+1}^T \dots \times_d \mat{A}_d^T$ is computed, 
we need to permute and reshape the tensor $\ten{B}$ into $\mat{B}^{(k)}$ before performing an SVD. Since $\ten{B}$ is stored in an external DRAM and the matrix $\mat{B}^{(k)}$ is stored in an on-chip memory, we need an extra module to move the data between the DRAM and the on-chip memory and re-organize the data elements to match $\mat{B}^{(k)}$. 
After SVD, the factor matrix need to be transposed and moved from the on-chip memory to DRAM, which is done by this module as well.

When moving the data from on-chip memory and external DRAM, the data is first read from the its original memory, written to a local buffer with size $p'\times q'$, then written to the destination memory. The size of the buffer determines the size of data set that can be moved in every batch.


\section{Implementation Details} 
\subsection{Fixed Point Number}
Floating-point numbers usually cause higher hardware cost and longer latency. Therefore, we use a fixed-point number system based on the trade-off between accuracy and hardware complexity.

We decide the fixed-point precision based on some hardware constrains. Because the data width at the interface of a DRAM controller is fixed as 512 bits, the memory is most efficient if the data width is a factor of 512 (i.e., $2^n$ with integer $n\leq 9$). Meanwhile, each DSP our FPGA can perform an multiplication with data sizes up to 27 bits $\times$ 18 bits. Considering these constraints, we use 16-bit numbers to represent all tensor data elements, and store them in an external DRAM. On the other hand, we use 27-bit numbers to represent the matrix data in both TTM and SVD in order to achieve higher accuracy and to avoid excessive use of multipliers. Note that we use a smaller data width for tensor data in order to save some memory space when processing the huge amount of data in a tensor. In this case, each multiplier in the TTM unit requires one DSP block, and each multiplier in the SVD unit requires two DSP blocks.

There are many sum-of-product operations in both TTM and SVD. The small error in the product terms will accumulate when calculating the sum. In order to address this issue, we use 48-bit numbers to represent the product terms. We use 27-bit numbers to represent most of the intermediate results, except for the 32-bit $\alpha, \beta, \gamma, \theta$ in the SVD unit.

\begin{algorithm}[t]
\caption{HOOI with warm-start Jacobi iterations}
\label{alg:warm}
\begin{algorithmic}[1]
\STATE {Initialize $\mat{A}_k$ as any orthonormal matrix.}
\WHILE{Not converge}
    \FOR {$k=1,2, \ldots, d$} 
        \STATE {$\ten{B} \leftarrow \ten{X} \times_1 \mat{A}_1^T \dots \times_{k-1} \mat{A}_{k-1}^T \times_{k+1} \mat{A}_{k+1}^T \dots \times_d \mat{A}_d^T$ }
        \STATE {Unfold $\ten{B}$ into $\mat{B}^{(k)}$}
        \STATE {SVD: run Jacobi iterations (i.e., Alg.~\ref{alg:jacobi}) with input $\mat{B}=\mat{U}_k^T \mat{B}^{(k)}$ and initial guess $\mat{U} = \mat{U}_k$.}
        \STATE {$\mat{A}_k \leftarrow$ the first $R_k$ columns of $\mat{U}$.}
    \ENDFOR
\ENDWHILE
\end{algorithmic}
\end{algorithm}

\subsection{HOOI with A Warm-start Algorithm}

We observe that the SVD $\mat{B}^{(k)}=\mat{U}_k\mat{S}_k\mat{V}_k$ via the Jacoboi iterations is the most time-consuming part of HOOI. Therefore, we employ a warm-start strategy to reduce the number of Jacobi iterations. In the standard Jacobi iterations, the initial guess is chosen as an identity matrix. In our implementation, we use the orthonormal matrix $\mat{U}_k$ obtained from the previous iteration of HOOI as the initial guess of the Jacobi iteration. Thanks to the good initial guess, we only need to perform one or two rounds of Jacobi iterations inside each SVD, and the whole HOOI still converges after an enough number of power iterations. The optimized algorithm is shown in Alg.~\ref{alg:warm}.

\begin{figure}[t]
    \centering
    \includegraphics[width=0.44\textwidth]{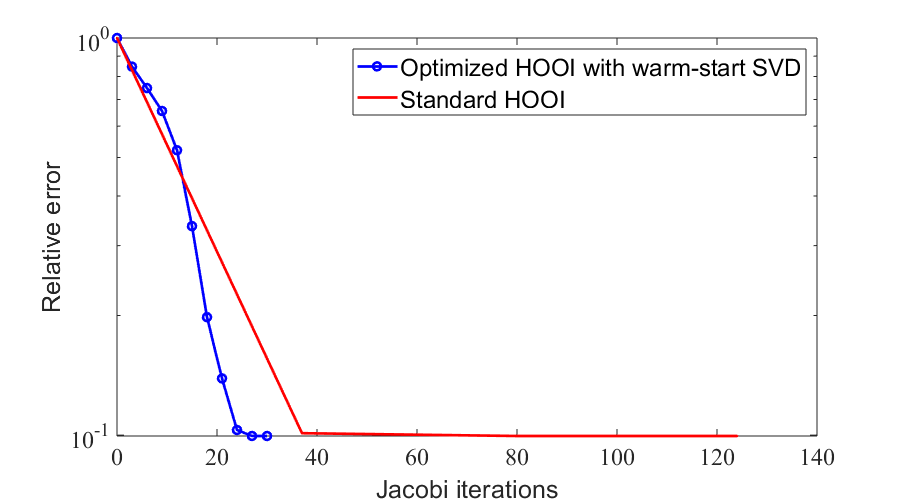}
    \caption{Convergence speed (measured as the total number of Jacobi iterations) of a standard HOOI and the optimized HOOI with the warm-start inside SVD.}
    \label{fig:iter}
\end{figure}

Fig.~\ref{fig:iter} shows the convergence curves of the standard HOOI and our proposed warm-start HOOI, respectively. We consider a tensor of size $128\times 128 \times 128$ and with a multi-linear rank $\mat{R}=[32,32, 32]$, which is generated by a Gaussian distribution and corrupted with some Gaussian noise. The noise variance is half of that of the tensor element. The standard method converges after only two HOOI iterations, but the SVD of each mode requires about 10 rounds of Jacobi iterations. Our optimized HOOI converges after 7 iterations, but each SVD requires only one round of Jacobi iterations, leading to a significant reduction of the total cost.

\section{Performance Analysis}
This section analyzes the hardware performance of our FPGA-accelerated Tucker decomposition.

\subsection{Run-time Analysis}
The total runtime is the sum of each part: TTM, SVD, and tensor permuting. For a $d$-way tensor, each HOOI iteration requires $d(d-1)$ TTMs, $d$ SVDs and $2d$ tensor permuting/reshaping operations. Some intermediate results can be reused. For instance, after computing $\ten{X}\times_3\mat{A}_3\times_2\mat{A}_2$ for a 3-way $\ten{X}$, the result of $\ten{X}\times_3\mat{A}_3$ can be reused when computing $\ten{X}\times_3\mat{A}_3\times_1\mat{A}_1$. By considering the TTM data reuse, the actual total number of TTMs is reduced to $(d-1)(1+d/2)$. When applying the warm-start algorithm for Jacobi iteration, the unfolded matrix $\mat{B}^{(k)}$ need to be multiplied with $\mat{U}_k$ first, and this is done by TTM as well, causing additional $d$ TTM operations. The total run-time is given by 
\begin{equation}
    T_{\text{total}}= \sum_{k=1}^{d}(\sum_{j=1}^{k}T_{\text{TTM}, k, j} + T_{\text{SVD}, k} + \sum_{i=1}^{2} T_{\text{permute}, i, k}).
\end{equation}
The details of each term are provided below.

\vspace{5pt}
\textbf{TTM Part:} The run-time of TTM depends on the size of its multiplier matrix. Suppose that we have a multiplier matrix of size $q \times r$. In this case, the multiplier takes in $q$ elements of $\ten{X}'$ per clock cycle, and each element is multiplied with $r$ elements of the factor matrix $\mat{A}_j$. At most $qr$ product and sum operations can be done per clock cycle. Therefore, the number of clock cycles is
{\small
\begin{equation}
T_{\text{TTM}, k,j}=\left\{
\begin{array}{ll}
I_j\prod \limits_{j'=j+1}^d R_{j'} \times \lceil { \frac{\prod\limits_{j'=1}^{j-1} I_{j'}}{q} }\rceil \times \lceil \frac{R_j}{r} \rceil   &  j>k \\
I_jI_k\prod\limits_{j'=j+1, j'\neq k}^d R_{j'} \times \lceil { \frac{\prod \limits_{j'=1}^{j-1} I_{j'}}{q} }\rceil \times \lceil \frac{R_j}{r} \rceil   & 1<j<k\\
I_k\prod \limits_{ j' =2,j' \neq k}^d R_{j'} \times \lceil \frac{I_1}{q} \rceil \times \lceil \frac{R_1}{r} \rceil & j=1.
\end{array} \nonumber
\right.
\end{equation}
}
Similarly, the clock cycles required for $\mat{U}^{(k)}\mat{B}_k $ is 
{\small
\begin{equation}
T_{\text{TTM}, k,k}=\left\{
\begin{array}{ll}
I_k\prod \limits_{j'=k+1}^d R_{j'} \times \lceil { \frac{\prod \limits_{j'=1}^{k-1} R_{j'}}{q} }\rceil \times \lceil \frac{I_k}{r} \rceil   &  k\neq 1\\
\prod \limits_{ j'=2}^d R_{j'} \times \lceil \frac{I_1}{q} \rceil \times \lceil \frac{I_1}{r} \rceil & k=1.
\end{array} \nonumber
\right.
\end{equation}
}

Some extra clock cycles are caused by the latency of the pipeline and ping-pong buffer, but they are often less than $1\%$ of the total run-time and thus negligible. When the TTM is applied over the first mode, the data need to be copied to the local memory of each PE in advance. This causes extra $O(I_k  \lceil \frac{J_k}{r} \rceil)$ clock cycles, which is negligible again and can be done in parallel with other operations.

\vspace{5pt}
{\bf SVD Part:} 
When updating the $k$-th factor matrix, we need to do SVD to a matrix with size $I_k \times R_{/k}$, with $R_{/k} = \prod_{i\neq k} R_i$. In each Jacobi iteration, $I_k(I_k-1)/2$ pairs of matrices will be orthogonalized. Each orthogonalization handles $2(I_k+R_{/k})$ numbers, therefore it takes $2\lceil \frac{I_k+R_{/k}}{p} \rceil$ clock cycles, where $p$ is the degree of parallelism. As a result, each Jacobi iteration takes approximately 
\begin{equation}
    T_{\text{SVD},k}=I_k(I_k-1)\lceil \frac{I_k+R_{/k}}{p}\rceil
    \label{eq:time_svd}
\end{equation} 
clock cycles. Similar to the case in TTM, the extra time cased by the latency of pipeline is negligible.

\begin{figure*}[t]
\centering
    \begin{subfigure}[b]{0.24\textwidth}
        \centering
        \includegraphics[width=\textwidth]{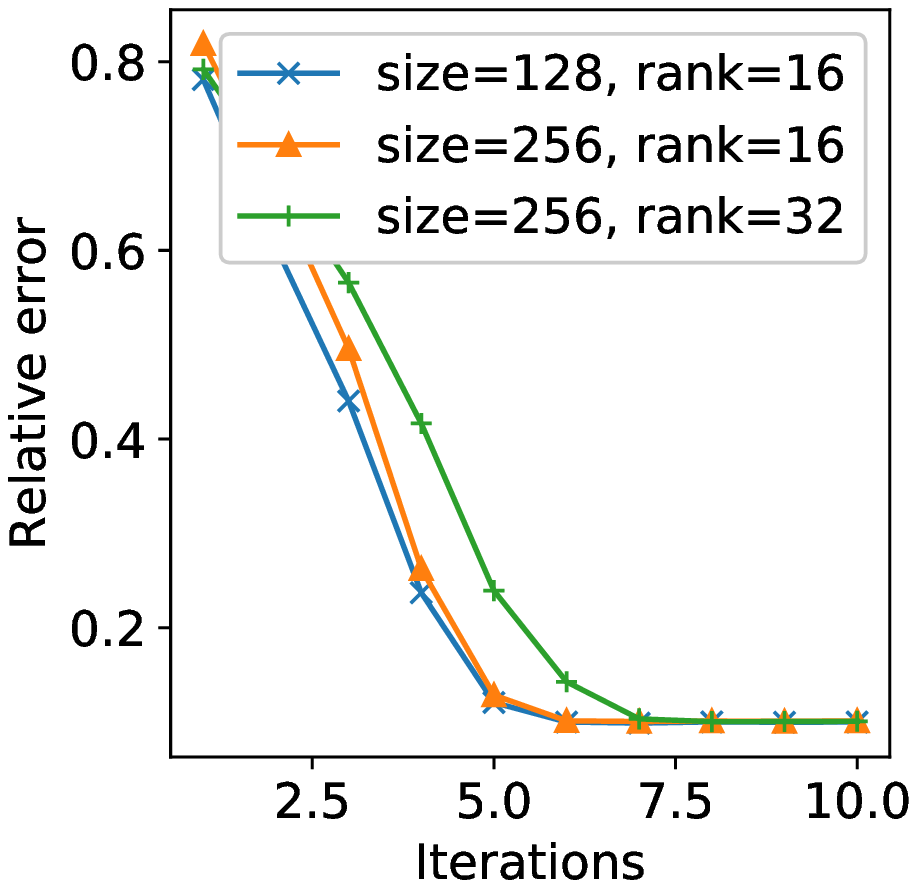}
        \caption{}
    \end{subfigure}
    \begin{subfigure}[b]{0.24\textwidth}
        \centering
        \includegraphics[width=\textwidth]{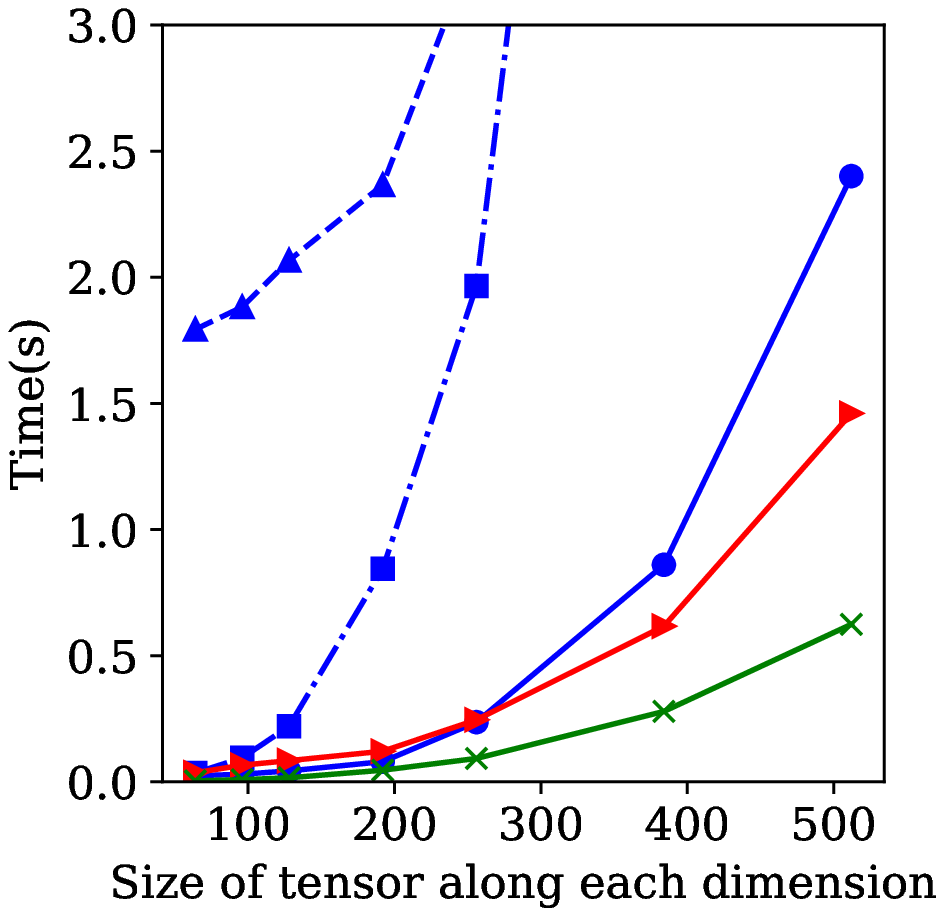}
        \caption{}
    \end{subfigure}
    \begin{subfigure}[b]{0.24\textwidth}
        \centering
        \includegraphics[width=\textwidth]{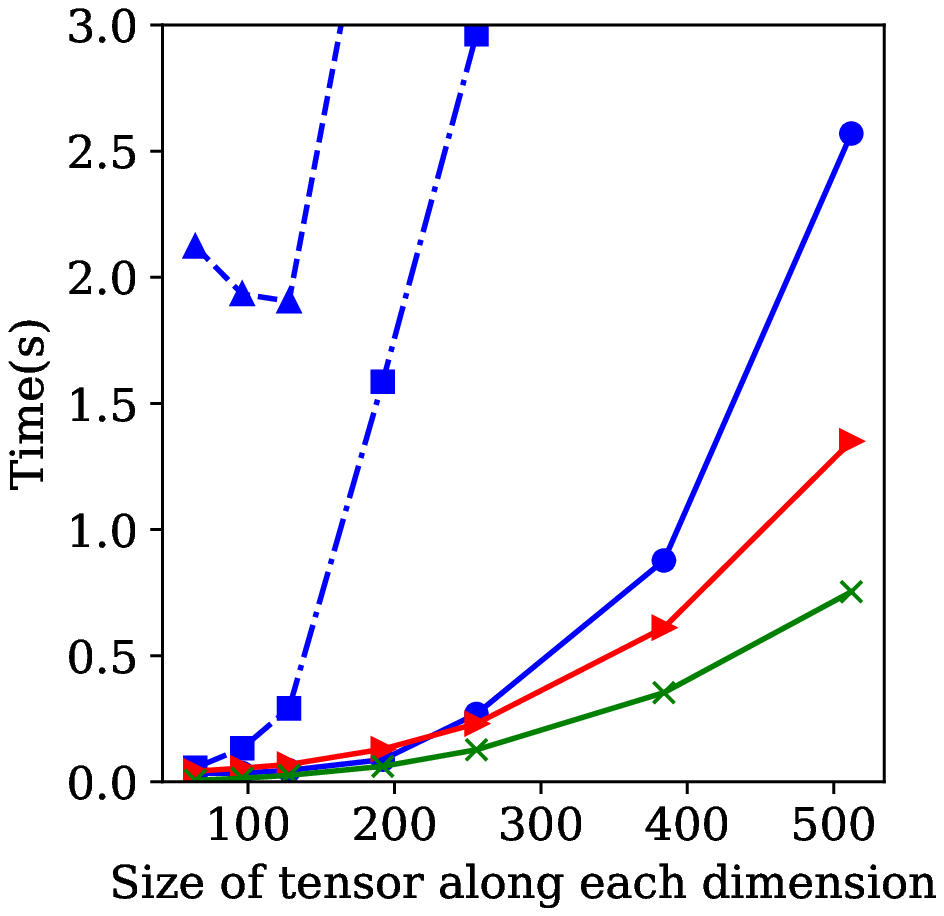}
        \caption{}
    \end{subfigure}
    \begin{subfigure}[b]{0.24\textwidth}
        \centering
        \includegraphics[width=\textwidth]{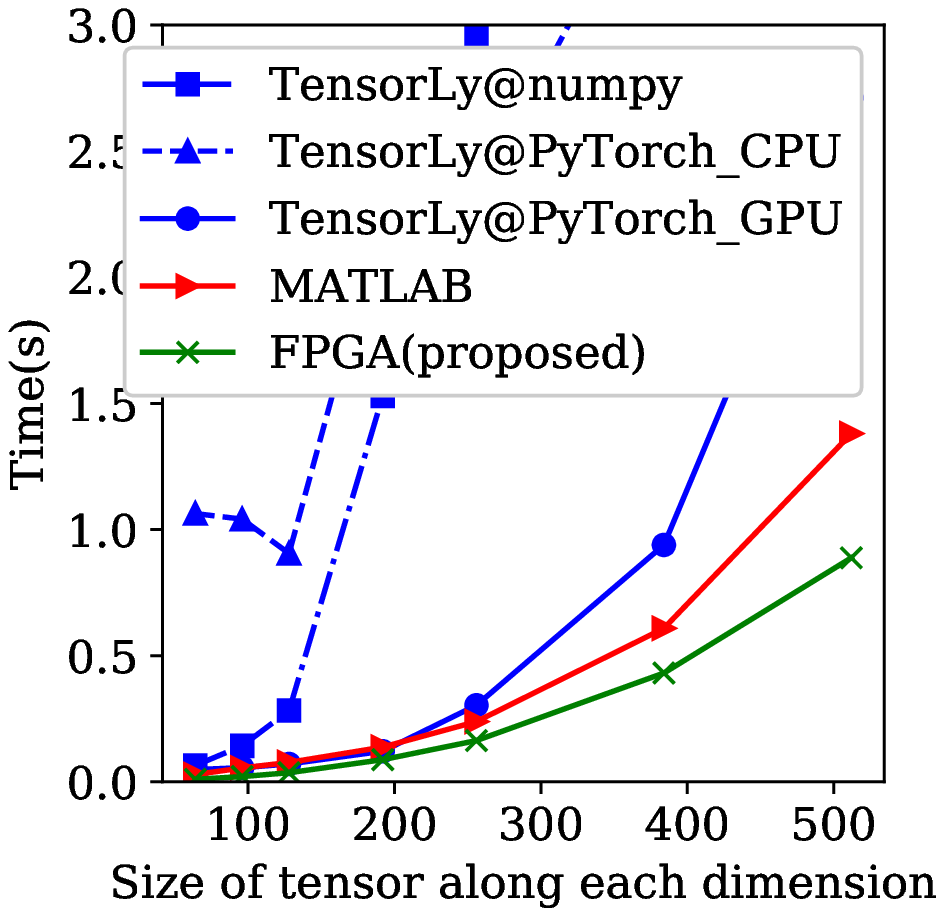}
        \caption{}
    \end{subfigure}
    
    \caption{Runtime and convergence of HOOI on some randomly generated 3-way tensors with size $I_1=I_2=I_3$ and $R_1=16, R_2=24, R_3=32$. (a) Convergence of proposed FPGA-based HOOI on a 3-way tensor, in the number of HOOI iterations. (b)-(d) Runtime of HOOI for various sizes of tensors. TensorLy uses the standard HOOI which uses SVD for initialization and for updating $\mat{A}_k$; the MATLAB implementation uses random initialization and eigenvalue decomposition to update $\mat{A}_k$; our proposed FPGA implementation uses random initialization and SVD Jacobi method with warmstart to update $\mat{A}_k$. }
    \label{fig:syn}
\end{figure*}

\begin{figure}[t]
\centering
    \begin{subfigure}[b]{0.22\textwidth}
        \centering
        \includegraphics[width=\textwidth]{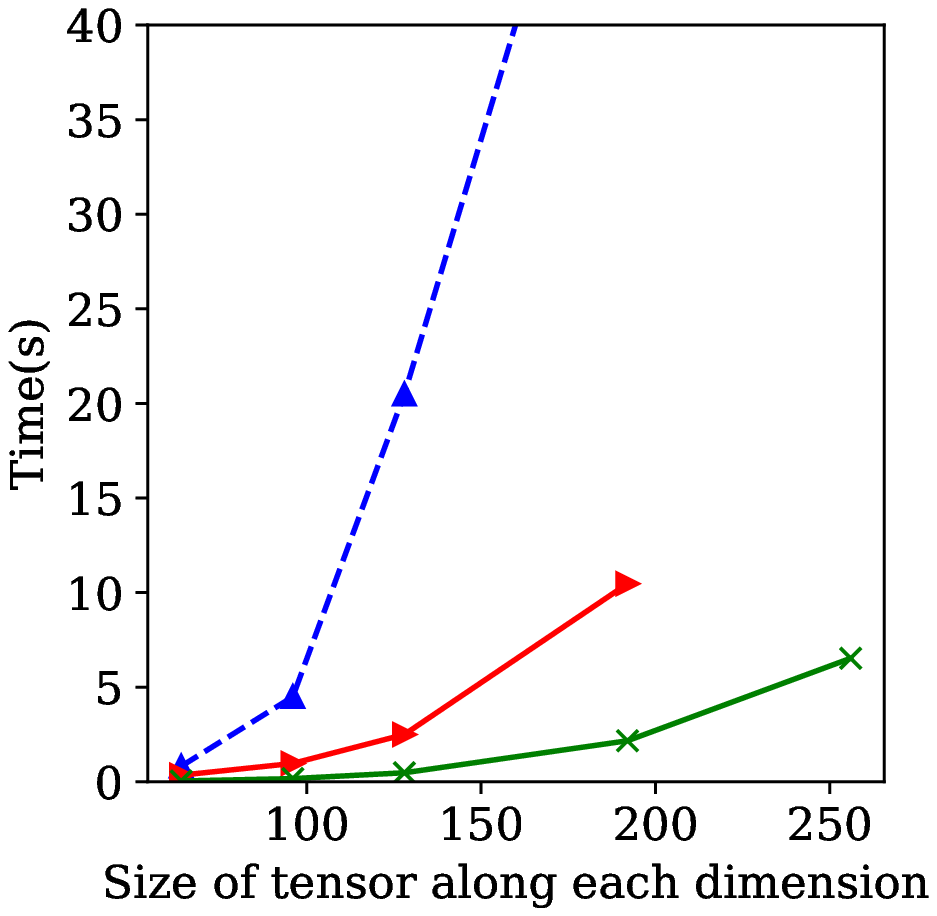}
        \caption{}
    \end{subfigure}
    \begin{subfigure}[b]{0.22\textwidth}
        \centering
        \includegraphics[width=\textwidth]{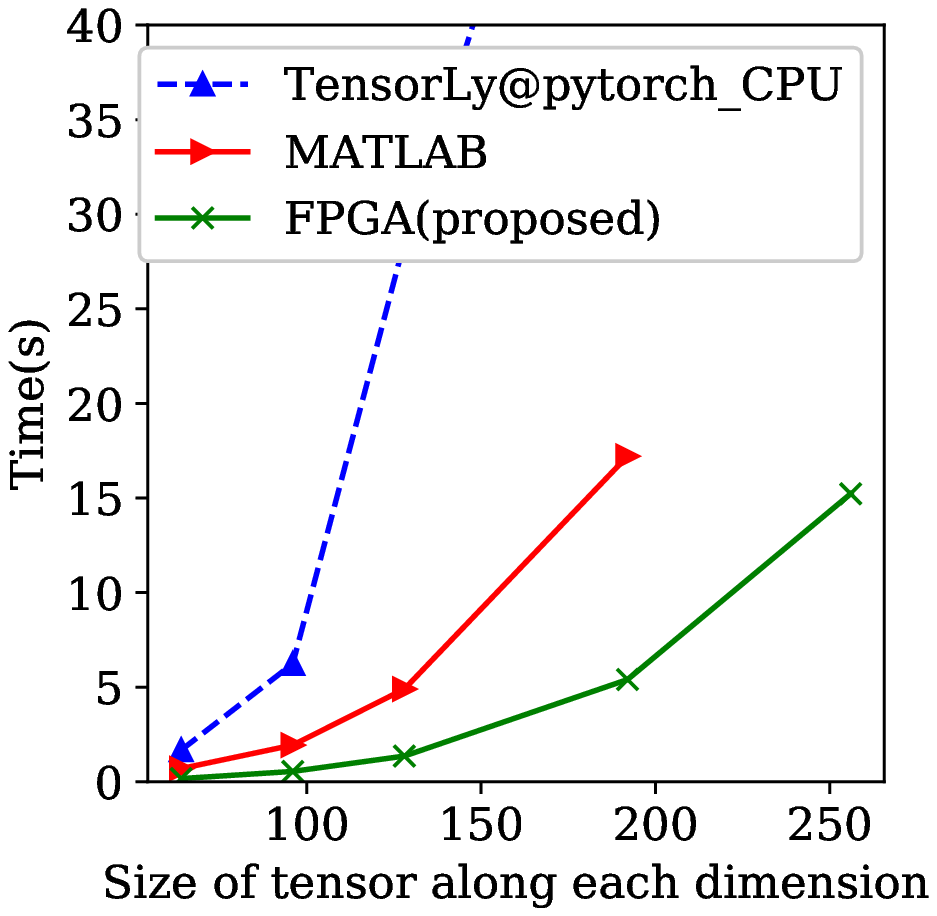}
        \caption{}
    \end{subfigure}
    \caption{Runtime of HOOI on some 4-way tensors with the same size along each dimension, and rank $\mat{R}=[16,16,16,16]$ and $\mat{R}=[32,32,32,32]$, respectively.}
    \label{fig:syn_2}
\end{figure}

\vspace{5pt}
{\bf Tensor Permutation:} Suppose that the size of the buffer is $p' \times q'$. In each clock cycle, this buffer can either exchange (read or write) $p'$ numbers with the internal memory or $q'$ numbers with the DRAM. Note that $p'$ and $q'$ are not necessarily equal to $p$ and $q$, respectively, as long as the memory supports writing $p'$ or $q'$ elements each time. However, setting $p'=p, q'=q$ can simplify our design and maximize the hardware efficiency. Each tensor permutation requires $O\left(I_k\lceil \frac{R_{/k}}{q'} \rceil+R_{/k}\lceil \frac{I_k}{p'} \rceil\right)$ clock cycles, with $R_{/k} = \prod_{i\neq k} R_i$. Our simulation shows that the practical run-time is 
\begin{equation}
    T_{\text{permute}, 1, k} \approx 5(I_k\lceil \frac{R_{/k}}{q'} \rceil+R_{/k}\lceil \frac{I_k}{p'} \rceil)
\end{equation} clock cycles when we move $\ten{B}$ from a DRAM to an on-chip memory and permute it. We need  \begin{equation}
    T_{\text{permute},2, k} \approx 5I_k(\lceil \frac{I_k}{q'} \rceil+\lceil \frac{I_k}{p'} \rceil)
\end{equation} clock cycles to move $\mat{U}_k$ to DRAM and transpose it. 


\subsection{Area and Power}

The area and power depends on the design parameters $p$, $q$ and $r$. The higher the degree of parallelism is, the more PEs, hardware area and power will be required. In the TTM unit, the total number of multipliers, adders, accumulators and buffers are proportional to the size of multiplier matrix, $q \times r$. Therefore, the area of TTM is approximately $O(q \times r)$. Similarly, the area of the tensor permutation unit is proportional to the buffer size $p \times q$. The power also increases when the area increases.

The area of the SVD unit is independent of its input matrix size, but depends on the degree of parallelism $p$. Additionally, one $\textbf{arctan}$ module and one $\textbf{sin/cos}$ module are required. Therefore, the area of the SVD unit is estimated as $c_1p+c_2$, where $c_1$ represents the area (multipliers, adders, accumulators, etc.) required for each matrix element, and $c_2$ represents the area of $\textbf{arctan}$ and $\textbf{sin/cos}$ blocks.

\section{Results}
\begin{table}[t]
    \centering
    \small
    \begin{tabular}{cc|ccccc}
        \hline
        $q$ & $r$ &  LUTs & Registers & DSPs & \shortstack{Clock\\rate} & Power \\
        \hline
        16 & 16 & 46,056  & 24,556 & 256 & 212MHz & 2.008W \\
        16 & 32 & 99,384  & 48,357 & 512 & 200MHz & 2.395W \\
        32 & 16 & 99,505  & 48,189 & 512 & 203MHz & 2.503W \\
        32 & 32 & 198,269 & 95,743 & 1,024 & 187MHz & 3.141W\\
        \hline
    \end{tabular}\normalsize
    \caption{Performance of the TTM unit.}
    \label{tab:ttm}
\end{table}
\begin{table}[t]
    \centering
    \small
    \begin{tabular}{c|ccccc}
        \hline
        $p$ &  LUTs & Registers & DSPs & Clock rate & Power \\
        \hline
        16  &  8,711 & 12,284 & 128 & 209MHz & 0.477W \\
        32  & 11,134 & 13,965 & 256 & 207MHz & 0.683W \\
        64  & 16,127 & 17,532 & 512 & 208MHz & 1.095W \\
        128 & 25,360 & 24,631 & 1,024 & 203MHz & 1.871W\\
        \hline
    \end{tabular}
    \normalsize
    \caption{Performance of the SVD unit (fixed point).}
    \label{tab:svd}
\end{table}
\begin{table}[!t]
    \centering
    \small
    \begin{tabular}{cc|cccc}
        \hline
        $q$ & $p$ &  LUTs & Registers & \shortstack{Clock\\rate} & Power \\
        \hline
        16 & 64  & 29,929  & 27,718 & 241MHz & 1.342W \\
        32 & 64  & 59,308  & 55,366 & 209MHz & 1.961W\\
        32 & 128 & 115,749 & 110,662 & 205MHz & 2.981W\\
        \hline
    \end{tabular}
    \normalsize
    \caption{Performance of the tensor permute/reshape unit.}
    \label{tab:per}
\end{table}

\subsection{FPGA Performance Validation}

We implement all parts of the optimized HOOI with different design parameters on FPGA. All the results below are based on Xilinx XCVU9P-L2FSGD2104E FPGA, which is available on a Xilinx VCU1525 acceleration board. The power is estimated with a 200-MHz clock rate. The results for different units are shown in Tables \ref{tab:ttm}-\ref{tab:per}. The hardware complexity, including the number of lookup tables (LUTs), registers and DSP blocks, is determined by the design parameters. The area of the TTM unit is approximately proportional to $q\times r$. The area of the SVD unit is approximately proportional to $p$. In tensor permute unit, its area is approximately proportional to $p'\times q'$. The power consumption increases as we increase the design parameters but the relationship is not strictly linear, since the power consumption of some parts (e.g., the clock generator) is independent of our design parameters. 
\subsection{Performance Comparisons}
We compare our FPGA accelerator with other two commonly used toolboxes on different platforms: the Tensor toolbox \cite{TTB_Software, TTB_Dense} in MATLAB on CPU, and the TensorLy toolbox \cite{tensorly} in Python on both CPU and GPU. The TensorLy can select NumPy, PyTorch or MXNet as its backend, and the latter two allow high-performance GPU computation. In our experiment, we use NumPy and PyTorch for comparison. The runtime on CPU is measured on a Linux desktop with Intel i5-8400 6-core CPU and 32 GB memory. The GPU experiments are conducted on a Titan V GPU. Since the accuracy and convergence of our Tucker decomposition depends on the fixed-point number system, we develop a fixed-point simulator with the Xilinx fixed-point number library to simulate the truncation error and overflow in our FPGA accelerator. 

We perform the comparison by using some randomly generated low-rank tensors. For each tensor, both the core tensor and the factor matrices are generated by Gaussian distributions, and the tensor is then corrupted by some Gaussian random noise. The relative error is evaluated as
\begin{equation}
    \frac{\|\ten{X} - \ten{G} \times_1 \mat{A}_1 \times_2 \mat{A}_2 \times \dots \times_d \mat{A}_d \|_{\rm F}}{\|\ten{A}\|_{\rm F}} \times 100\%.
\end{equation}

Fig. \ref{fig:syn} shows the results on a set of 3-way tensors with size $I_1=I_2=I_3$. Our architecture can get $1.41-9.90\times$ speedup compared with MATLAB tensor toolbox on CPU, and even more compared with the TensorLy toolbox on both CPU and GPU.
The convergence behavior of our FPGA-based Tucker decomposition is shown in Fig.~\ref{fig:syn}(a). It is shown that our method always converges after 6-8 HOOI iterations. Therefore, we assume 8 HOOI iterations to estimate the runtime of our FPGA architecture. 

We further perform comparisons using some 4-way tensors and show the results in Fig.~\ref{fig:syn_2}. Our PC with 32GB RAM can no longer accommodate such 4-way tensor data, therefore the results on CPU are obtained by running our experiments on Amazon AWS r4.4x large instance with 16 virtual CPUs (vCPU) and 122GB memory. Since large 4-way tensors exceed the memory of GPU, and TensorLy with the NumPy backend consumes extremely long time, their runtimes are not shown in Fig.~\ref{fig:syn_2}. When the size along each dimension is 256, MATLAB run out of memory when computing. On these 4-way tensor data, our FPGA design can get $3.18-8.22\times$ speedup compared with the MATLAB tensor toolbox on the AWS CPU.

{\bf Remark:} Our FPGA accelerator uses the Jacobi iteration to perform SVD, whereas the MATLAB tensor toolbox and TensorLy use more powerful advanced SVD algorithms. If the same SVD algorithms are used in all implementations, our FPGA design may achieve more significant speedup.

\subsection{Application: MRI Compression}

\begin{figure}[t]
\centering
\includegraphics[trim=0 0.5cm 0 1cm,clip, width=0.10\textwidth]{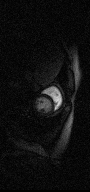}
\includegraphics[trim=0 0.5cm 0 1cm,clip, width=0.10\textwidth]{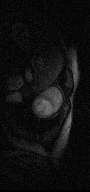}
\includegraphics[trim=0 0.5cm 0 1cm,clip, width=0.10\textwidth]{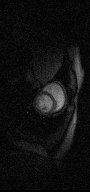}
\includegraphics[trim=0 0.5cm 0 1cm,clip, width=0.10\textwidth]{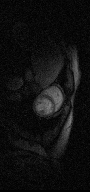}\\ \vspace{0.05in}
\includegraphics[trim=0 0.5cm 0 1cm,clip, width=0.10\textwidth]{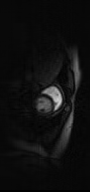}
\includegraphics[trim=0 0.5cm 0 1cm,clip, width=0.10\textwidth]{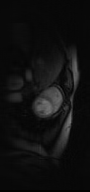}
\includegraphics[trim=0 0.5cm 0 1cm,clip, width=0.10\textwidth]{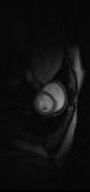}
\includegraphics[trim=0 0.5cm 0 1cm,clip, width=0.10\textwidth]{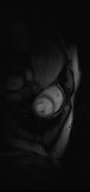}
\caption{Decomposition result of MRI dataset. Top: original $190\times 90\times70$ data at $t=15, 30, 45, 60$. Bottom: approximated data by Tucker decomposition on FPGA, with rank $\mat{R}=[40,32,28]$. The data compression ratio is $24.8 \times$.}
\label{fig:MRI}
\end{figure}

The accelerated Tucker decomposition can be applied to multiple application domains. Here we demonstrate its application in compressing multi-way MRI data. This dataset is a single-channel, first-pass myocardial perfusion real-time MRI sequence ($n_x = 190, n_y = 90, n_t = 70$). We use the pre-processed data available in \cite{sure_svt, lingala2011accelerated}, and set rank to $R_x=40,  R_y = 32, R_t = 28$ in HOOI. The estimated runtime with our architecture is 0.0447s at a clock rate of 185MHz. In comparison, on a Linux PC with Intel i5-8500 CPU 6 core CPU, the same operation takes 0.0964s with the MATLAB tensor toolbox, 0.335s with TensorLy toolbox and NumPy backend, 1.352s with TensorLy toolbox and Pytorch backend. on a PC with NVIDIA TITAN V GPU, this operation takes 0.217s. Therefore, our method provides 2.16-30.2$\times$ speedup compared with existing frameworks on CPU, and 4.85$\times$ speedup compared with GPU. Fig.~\ref{fig:MRI} shows the original MRI data and the data approximated by our low-rank Tucker decomposition on FPGA. 

\section{Conclusion}
\label{sec:conclusion}
This paper has presented an algorithm-architecture co-design to perform tensor Tucker decomposition. We have implemented Tensor-Times-Matrix, matrix SVD with single side Jacobi iteration, and tensor permuting on FPGA. We have also proposed a warm-start algorithm for the Jacobi iterations to reduce its computation time. We have done simulations on both synthetic data sets and an MRI data set. Our method is significantly faster than existing computation frameworks on both CPU and GPU. This accelerator can be employed in a broad range of applications including data mining, scientific computing, computer vision, and deep learning.

\section*{Acknowledgment}
This work was supported by NSF CCF 1817037. We would like to thank Sophia Shao (NVIDIA), Lei Deng, Zheng Qu and Bangyan Wang (UCSB) for their technical discussions.

\balance

\bibliographystyle{IEEEtran}
\small{
\bibliography{RefList}
}

\end{document}